\documentclass[]{spie}
\usepackage[]{graphicx}

\title{
Method of eigenvalue transformation using dispersion oscillating fiber
}

\author{Alexey A. Sysoliatin\supit{a}, Andrey I. Konyukhov\supit{b}
\skiplinehalf
\supit{a}Prokhorov General Physics Institute of RAS, 119991, Moscow, Vavilov Str.38, Russia; \\
\supit{b}Saratov State University, Astrakhanskaya 83, 410012, Saratov, Russia
}

\authorinfo{Further author information: (Send correspondence to A.Sysoliatin)\\A.Sysoliatin: E-mail: alexs@fo.gpi.ru, Telephone: +7 (499)503 87 02\\  A.Konyukhov: E-mail: kai@optis.sgu.ru, Telephone: +7 (8452)210 728}

\begin{document} 
\maketitle

\begin{abstract}
It is known that eigenvalues of Zakharov-Shabat problem  
can be used to encode a signal in soliton communication lines.
We propose to use dispersion oscillation
fiber as a new versalite tool to control the eigenvalues. A fiber
with sine-wave variation of the core diameter can be used to 
manage both real and imaginary parts of the
eigenvalues. Change of real part of the eigenvalues results in a splitting
of an optical breather into two distinct pulses propagating with the
different group velocities. Change of imaginary parts of the eigenvalues
allows to realize a reverse process of merge of two solitons into
high-intensity pulse. The splitting of optical breather and merge
of solitons can be obtained even under the strong effect of stimulated
Raman scattering. We believe tools and techniques based on use of
dispersion oscillating fiber will grant unprecedented control over soliton
eigenstates.
\end{abstract}


\section{INTRODUCTION}

Optical solitons are formal solutions of nonlinear Schr\"odinger equation
(NLSE) \cite{AkhmedievBook, Agrawal, Akhmanov}. Using so-called nonlinear
Fourier transform \cite{Yousefi}, a unique set of soliton
eigenvalues can be found. Control of soliton eigenvalues has practical
application in
all-optical routing \cite{AkhmedievBook}, 
nonlinear frequency division multiplexing \cite{Turitsyn, Le, Hasegava3}, 
controlled rogue wave generation \cite{Bendahmane, Weerasekara} and in generation of photon
entangled pairs \cite{SolitonCoding, Kivshar}.

The discrete eigenvalues remain unchanged during 
interaction of classical Schr\"odinger solitons. A special 
type of nonlinearity, dispersion or local perturbation
can be used to control soliton eigenstates, soliton amplitudes and velocities \cite{AkhmedievBook}.
The optical solitons can break up, merge or even annihilate 
\cite{Malomed, 20, 21,23}.
The role of perturbation can be played by variation of the fiber dispersion or
nonlinearity \cite{Hasegava, Chao-Qing},  
cross-phase modulation in a system of coupled NLS equations \cite{Mihalache, Lu},
and effect of stimulated Raman scattering \cite{5, 6}.
Processing of optical pulses via solitons interaction
has been analyzed using various nonlinear evolution models \cite{2f,4f,7f,9f}.
Soliton fusion was observed experimentally \cite{Friberg}.
The problem of transforming one optical pulse into
another via nonlinear propagation in dispersion varying optical
fiber was considered in Ref.~\citenum{Broderick}.

The soliton splitting or soliton fission 
can be controlled by resonance effects.
If the perturbation modulation period
is comparable to the oscillation period of a multisoliton pulse,
it breaks up into several fundamental solitons \cite{Hasegava}. 
Optical solitons propagating in fibers with a diameter
varying along their length satisfy a nonautonomous NLSE
with variable dispersion and nonlinearity coefficients \cite{Serkin}.
In an NLSE model with a harmonic potential \cite{7q}, a periodic
variation in the potential leads to decay of coupled soliton
states. 
Yan and Dai \cite{5q} considered a generalized NLSE with variable
coefficients. With different choices of the coefficients, 
the cubic, periodic, and parabolic solitons can be obtained
\cite{Liu}.
Dispersion, nonlinearity and gain coefficients
varying in a certain way may lead to the formation of
a rogue wave, with a periodic potential as a trigger mechanism
\cite{6q}.

We consider the fission of optical breather and fusion of soliton pair which are 
initiated by the periodical variation of the fiber dispersion.
In our experiments an optical fiber with periodically
varying diameter breaks up two-soliton pulse into
two fundamental solitons. By changing the modulation of the fiber,
one can control the group velocity, centre frequency and peak
power of such solitons. Using numerical simulation,
we demonstrated that dispersion oscillating fiber can be used for
fusion of optical solitons as well. The inelastic soliton interactions driven by 
variation of the fiber dispersion allow to manage eigenvalues.
Our study introduces all-optical tool for processing optical solitons.

\section{Nonlinear Schr\"odinger equation model}

To describe the pulse propagation in dispersion oscillating optical fibers with
the anomalous dispersion and Kerr-type nonlinearity, one can
employ the nonlinear Schr\"odinger equation model with variable coefficients:

\begin{equation}
\displaystyle
\frac{\partial A}{\partial z} + \frac{\alpha}{2} A(z,t) =  
-\,i\frac{\beta_2 (z)}{2} \frac{\partial^2 A}{\partial t^2} + 
\frac{\beta _3 (z)}{6} \frac{\partial^3 A}{\partial t^3}
+i\left( {P_{NL} + i\frac{1}{\pi \nu_0 }
\frac{\partial P_{NL} }{\partial t}} \right),
\label{schroedinger}
\end{equation}

\noindent
where $A(z,t)$ is the complex pulse envelope, $z$ is the
propagation distance; $t$ is the local time in the coordinate system
$(z = z, t \to t - z/u)$ \cite{Agrawal}; $u$ is the group velocity of the pulse,
$\alpha$ is the loss coefficient, and $\nu_0$ 
is the carrier frequency of the pulse. 
Functions $\beta _{2}(z)$ and $\beta _{3}(z)$ describe dispersion variation
which arises due to variation of the fiber diameter.
From the measurements of the dispersion of three different fibers drawn from  
the same preform, but having different diameters, we found following approximations: 
\begin{equation}
\beta_{2,3}(z) = \langle \beta_{2,3} \rangle
[ {1+\beta_{2,3(m)}\;\sin (2\pi \;z / z_m +\varphi_m)}],
\label{beta2}
\end{equation}
\noindent
where
$\langle \beta_2 \rangle = -12.76\,\mbox{ps}^{2}\mbox{km}^{-1}$,
$\langle \beta_3 \rangle = 0.0761\,\mbox{ps}^{3}\mbox{km}^{-1}$,
$\beta_{2(m)} = 0.02$,
$\beta_{3(m)} = 0.095$.

Nonlinear media polarization in (\ref{schroedinger}) includes the Kerr effect and 
delayed Raman scattering 
$P_{NL}(z,t)=\gamma(z) [(1 - f_R) |A|^2 A + f_R Q A(z,t)]$,
where 
$\gamma(z) =\langle \gamma \rangle \left( 1 + 0.028 \sin(2\pi z/z_m) \right),$
$\langle \gamma  \rangle = 10\,\mbox{W}^{-1}\mbox{km}^{-1}$, $f_R = 0.18$ \cite{Agrawal}.
The Raman delayed response $Q(z, t)$ 
is approximated by damping oscillations associated
with a single vibration mode \cite{Akhmanov}:
\begin{equation}
\frac{\partial^2 Q}{\partial t^2} +
\frac{2}{T_2} \frac{\partial Q}{\partial t}+
\Omega^2 Q(z,t) =\Omega^2 |A(z,t)|^2,
\end{equation}
where $T_2=32\,\mbox{fs}$, $\Omega=13.1\,\mbox{THz}$. 

In our simulations we assume that fiber loss is compensated $(\alpha=0)$ 
using, for example,
Raman gain \cite{Raman_gain}.
For numerical simulations a split-step Fourier
method \cite{Sinkin} was used. To suppress
the waves reflected from the boundaries of the calculation
window, we use absorbing boundary conditions.

The eigenvalues of Zakharov-Shabat problem \cite{Yousefi} are associated with 
soliton solutions of classical NLSE \cite{Akhmanov}.
To obtain information about the soliton content of numerical solution 
of the NLSE with variable coefficients (\ref{schroedinger}) the 
modified inverse scattering method \cite{jetph} was used. 
Using numerical simulations we obtain eigenvalues as functions
of their propagation distance: $\lambda_j = \lambda_j(z)$. 
The complex envelope of the $j$-th fundamental soliton is defined as follows:
      \begin{equation}
      A_j(z,t )= R_j \, {\rm sech} (\kappa_j t - t_j - v_j z) \exp(i \varphi_j (z,t )) ,
      \label{As}
      \end{equation}

\noindent
where $R_j$ is the soliton amplitude;
$\kappa_j$ is the inverse soliton duration; $t_j$ is 
the coordinate of the pulse peak, $\varphi_j(z, t)$ is the field phase; 
$v_j$ determines the change
in the group velocity of the soliton: $v_g = (1/u + v_j)^{-1}$. 
The amplitude, duration and group velocity of the fundamental soliton (\ref{As}) can be
expressed through the spectral parameters $\lambda_j$:
      \begin{eqnarray}
      R_j &=&\tau_0^{-1} (|\beta_2|/\gamma)^{1/2} (2Im(\lambda_j)) ,\label{R} \nonumber \\
      \kappa_j &=&\tau_0^{-1} (2 Im(\lambda_j)) ,                   \label{k} \\
       v_j &=&\beta_2 \tau_0^{-1} (2 Re(\lambda_j)) ,               \label{v} \nonumber
\end{eqnarray}

\noindent
where $\tau_0 = t_0 (|\beta_2|/\gamma)(|\langle \beta_2 \rangle|/\langle \gamma \rangle)^{-1}$ 
is the duration of the fundamental soliton in a waveguide with adiabatic $(z_m >> 1)$
variations in the parameters $\beta_2(z)$ and $\gamma(z)$. 
Change of the soliton group velocity is associated 
with the shift of the pulse carrier frequency
\begin{equation} 
\Delta \Omega =2 \tau_0^{-1} Re(\lambda_j).
\label{frequency} 
\end{equation}

\section{Splitting of optical breather}

The nonlinear Schr\"odinger equation in absence of perturbations can support
bound states formed by multiple solitons localized at same positions.
Such states are known as breathers. For the breather
the input field can be described by hyperbolic secant pulse
\begin{equation}
A(0,t) = \frac{N}{t_0} \sqrt{\frac{|\beta_2|}{\gamma}} 
         {\rm sech}\left(\frac{t}{t_0} \right),
\label{sech}
\end{equation}
\noindent
where $A(0,t)$ is electric field envelope at the input ($z=0$),
$N$ is the soliton order, $t_0 = T_{_{\rm FWHM}}/1.76$ is the width 
of hyperbolic secant pulse,
$T_{_{\rm FWHM}}$ is the pulse width at the half of the
maximum intensity \cite{autocorrelation}, $\beta_2$ is the second-order
dispersion coefficient, $\gamma$ is the nonlinear coefficient \cite{Agrawal, Akhmanov}.
Note that $N$ defines the peak amplitude and generally is not integer number.
The breathers constitute a localized entity whose shape oscillates along 
propagation distance $z$ periodically. The oscillation period (soliton
period) \cite{Agrawal, Akhmanov,  Kivshar} is 
\begin{equation}
z_0 = \frac{\pi}{2} \frac{t_0^2}{|\beta_2|}
\label{z0}
\end{equation}

\begin{figure} [ht]
   \begin{center}
   \begin{tabular}{cc}
   (a)  & (b)  \\
   \includegraphics[height=4.5cm]{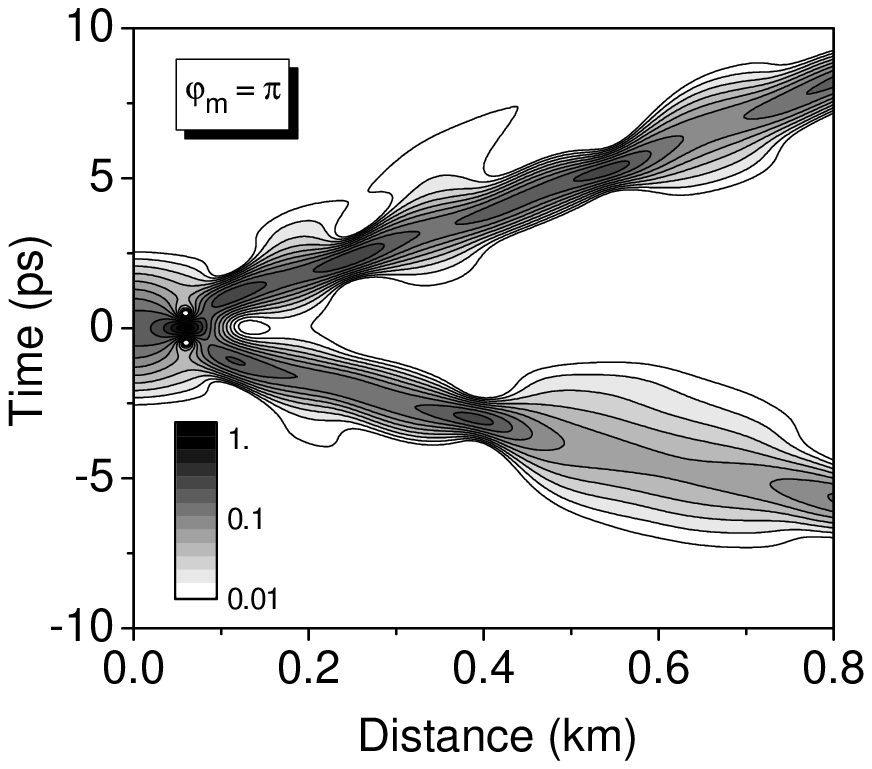} &
   \includegraphics[height=4.5cm]{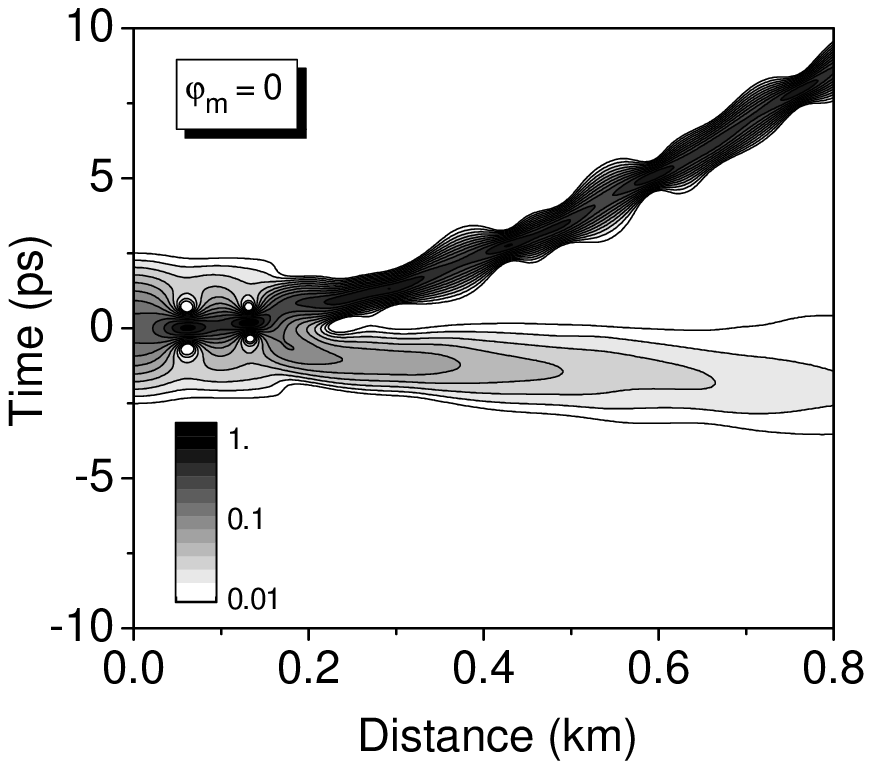}
   \end{tabular}
   \end{center}
   \caption[figure 1] 
   { \label{fig1} 
   Simulation of breather splitting into the pulse pair. 
   Contour plot of the field intensity. 
   (a) Dispersion modulation phase is $\varphi_m =\pi$.
   (b) Dispersion modulation phase is $\varphi_m = 0$.
   Energy of input pulse is 12.5~pJ. 
   Width of initial pulse (\ref{sech}) is $T_{_{\rm FWHM}}=2.1\,\mbox{ps}$;  
   Plots are normalized.  Colorbar indicates the range of intensity values.
   }
   \end{figure}

Using resonant modulation of the fiber dispersion we can split 
the second-order soliton into two pulses \cite{Sysoliatin}. 
Pulse dynamics obtained from numerical solution of  modified
nonlinear Schr\"odinger equation (\ref{schroedinger}) which includes
effect of stimulated Raman scattering, third-order dispersion and 
pulse self-steepening
in the Fig.\ref{fig1} is shown.

For dispersion modulation phase $\varphi_m = \pi$, only one modulation period of
DOF ($z=0.16\,\mbox{km}$) is sufficient for the splitting of initial breather (Fig.\ref{fig1}a). 
After $z=0.16\,\mbox{km}$ peak intensities of the pulses are reduced that
leads to reduce of the effect of stimulated Raman scattering in further pulse propagation.
For $\varphi_m = 0$,  the breather splits after propagation distance $z=0.24\,\mbox{km}$
(Fig.\ref{fig1}b). Increased propagation distance of high-intensity breather 
leads to enhance effect of stimulated Raman scattering. As results the 
output pulses have very different peak intensities.
Similar phenomena was observed experimentally \cite{Sysoliatin}. 

Eigenvalues of solitons which form breather (\ref{sech}) are given by formula \cite{Hasegava3}
\begin{equation}
\lambda_j = i (N - j + 1/2),
\label{lambdaIni}
\end{equation}
\noindent
where $j=1,2, \dots $ and $j<(N+1/2)$.
Imaginary part of the eigenvalue $Im(\lambda_j)$ give soliton amplitude (\ref{R}). 
Real part of the eigenvalue $Re(\lambda_j)$ gives shift of the soliton carrier 
frequency (\ref{frequency}).
This spectral shift in turn shifts the pulse position in the
time domain because of changes in the group velocity through fiber dispersion.

\begin{figure} [ht]
   \begin{center}
   \begin{tabular}{cc}
   (a) & (b) \\
   \includegraphics[height=4.cm]{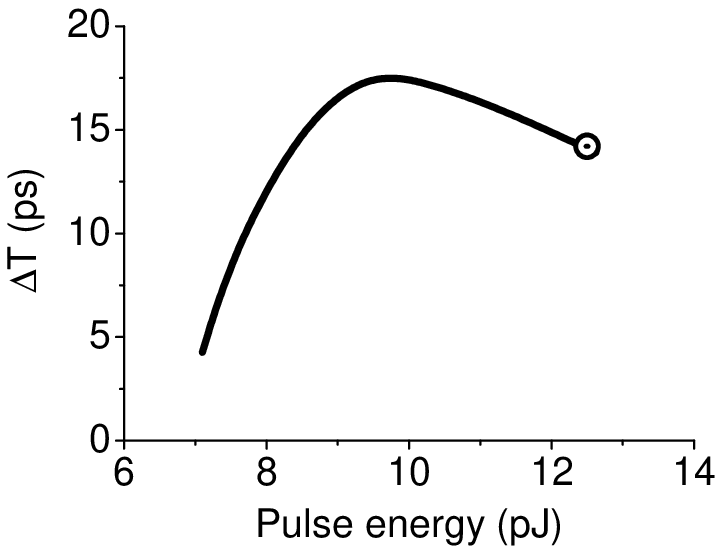} &
   \includegraphics[height=4.cm]{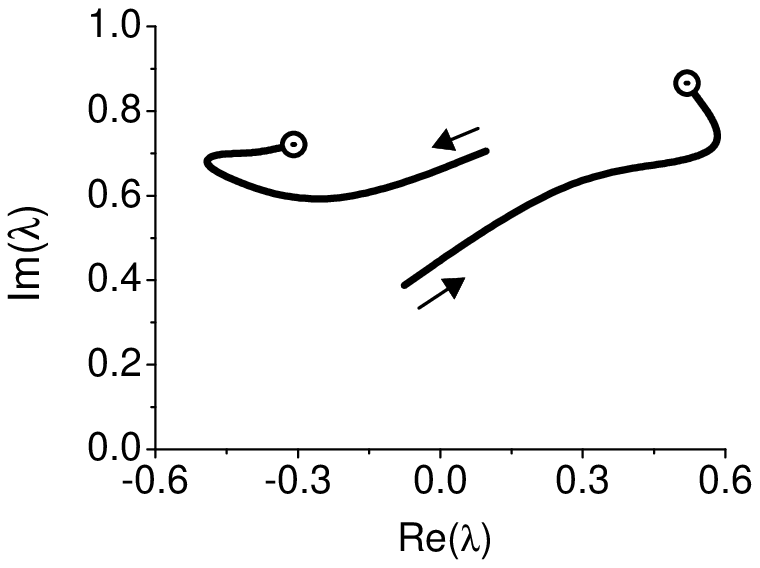} \\
   \end{tabular}
   \end{center}
   \caption[figure 2] 
   { \label{fig2} 
   Split of the breather in DOF. 
   (a) Calculated peak to peak distance for output pulses. Open circle shows the point,
   which corresponds to experimental result with 12.5~pJ input pulse energy; 
   (b) Movement of eigenvalues of output solitons with increase of the energy of input pulse 
   from 7.1~pJ to 12.5~pJ. Arrows show the moving direction of the eigenvalues. 
   Open circles corresponds to 12.5~pJ input pulse energy. 
   The results were obtained for $\varphi_m =\pi$ 
   Initial pulse width is $T_{_{\rm FWHM}}=2.1\,\mbox{ps}$.
   }
   \end{figure}

Eigenvalues (\ref{lambdaIni}) of initial solitons are purely imaginary.
That means all solitons in breather (\ref{sech}) propagates with the same
group velocity. The DOF changes soliton spectrum and real part of
eigenvalues becomes nonzero. As result the solitons which form initial
breather become separated in time. Figure \ref{fig2}a shows the 
change of peak to peak distance $\Delta T$ with
increase of the energy of input pulse from 7.1~pJ to 12.5~pJ.

Eigenvalues of two solitons that arize
due to the split of initial breather (\ref{sech}) in DOF are shown in
Fig.\ref{fig2}b. Curves in the Fig.\ref{fig2}b were calculated by increase
of the energy of input pulse. Directions of the movement of eigenvalues are
shown by arrows. Starting points correspond to 7.1~pJ input pulse energy
(soliton order is $N=1.65$). End points are marked by open circles and
correspond to 12.5~pJ input pulse energy ($N=2.2$). Figure \ref{fig2}c
shows that DOF can change $\lambda_j$ from purely imaginary values
(\ref{lambdaIni})  at input to the values with distinct real part. The
imaginary part of output solitons can be equalized by adjusting of input
pulse energy.

In model of unperturbed NLSE the second-order
breather ($N=2$) splits into solitons with eigenvalues which imaginary parts are equal
and real parts are opposite in sign $Re(\lambda_1) = -Re(\lambda_2)$.
This property follows from energy and momentum conservation laws \cite{Agrawal, Akhmanov}
for unperturbed Schr\"odinger equation.
The main process that disturbs the splitting of the breather into two pulses with
identical energies $Im(\lambda_1) = Im(\lambda_2)$
and opposite in sign frequency shift $Re(\lambda_1) = -Re(\lambda_2)$ is 
stimulated Raman scattering. 

The results described above give optical all-fiber tool for the change of
eigenvalues of picosecond solitons. The output pulses have different carrier 
frequencies that can be used to build up the high repetition rate 
multiwavelength optical clock.

\section{Inelastic two-soliton interaction}

Elastic collision of two solitons in case of ideal nonlinear
Schr\"odinger equation results only
in a shift of their phases and positions \cite{Agrawal, Akhmanov}.
However, even a relatively small perturbation changes this property.
Under perturbation the solitons can change their velocities and
amplitudes. In this section we describe inelastic soliton interactions
forced by periodical variation of the fiber dispersion.
As initial field a superposition of two single-soliton pulses was considered
\begin{equation}
A(0, t) = A_0 \,{\rm sech}(t / t_0 - T) + A_0 \,{\rm sech}(t / t_0 + T),
\label{Aini}
\end{equation}

\noindent
where $t_0 = 1.13$~ps is the initial pulse duration; 
$A_0 = t_0^{-1} \left( |\langle \beta_2 \rangle | / \langle \gamma \rangle \right)^{1/2}$ 
is the initial single-soliton pulse amplitude \cite{Akhmanov, Agrawal}.
The dimensionless parameter $T$ determines the separation
between the peaks of the initial pulses.
For the soliton pair (\ref{Aini}) propagating in a constant-diameter fiber $(z_m = \infty)$
there exists an analytical solution \cite{Akhmanov,  Kivshar, TMF}. 
In-phase solitons (\ref{Aini}) periodically attract and repel. 
Period of oscillations is given by
\begin{equation}
z_p = 2 z_0 ( \lambda_1^2 - \lambda_2^2)^{-1},
\label{z_p}
\end{equation}

\noindent
where $z_0=0.158~\mbox{km}$ (\ref{z0})
is the soliton period \cite{Agrawal,Akhmanov} 
in a constant diameter fiber $(z_m = \infty)$.
Complex numbers 
$\lambda_1=i0.5(\sinh(T)-1)/\cosh(T)$ and 
$\lambda_2=i0.5(\sinh(T)+1)/\cosh(T)$
are soliton eigenvalues \cite{TMF}.
For non-interacting pulses $(T=\infty)$ the period of oscillations $z_p$ tends to infinity
and soliton eigenvalues become equal $\lambda_1=\lambda_2=i0.5$.
In the case of the NLSE with constant coefficients $(z_m = \infty)$,
solitons interact elastically and their parameters $\lambda_j$ remain
unchanged \cite{Akhmanov, Agrawal, Le}. If there is dispersion modulation, 
interaction between solitons may have an inelastic nature, and 
the parameters $\lambda_j$ are variable. 

\begin{figure} [ht]
   \begin{center}
   \begin{tabular}{cc} 
   (a) & (b) \\
   \includegraphics[height=4cm]{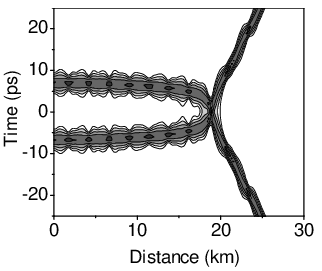} & 
   \includegraphics[height=4cm]{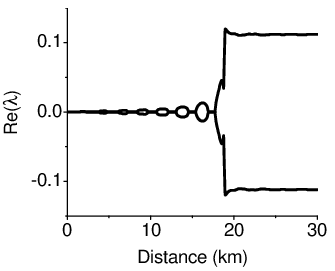} \\
   (c) & (d) \\
   \includegraphics[height=4cm]{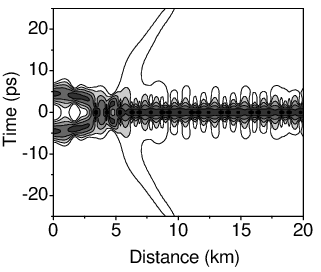} &
   \includegraphics[height=4cm]{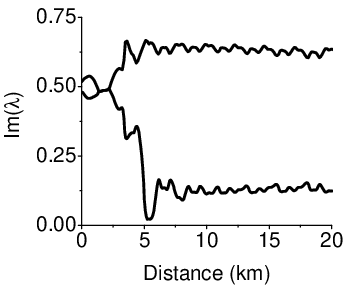}
   \end{tabular}
   \end{center}
   \caption[figure 4] 
   { \label{fig4} 
   Inelastic collision of two in-phase solitons in dispersion oscillating fiber
   (a,b) $T=6$;  (c,d) $T=4$;
   (a,c), Contour plots of the field intensity $|A(z,t)|^2$; 
   (b) Real part of eigenvalues vs. propagation distance.
   (d) Imaginary part of eigenvalues vs. propagation distance.
   Modulation period is $z_m =2.4\,\mbox{km}$, modulation phase is $\varphi_m =0$.}
   \end{figure}

The inelastic soliton collision in the Fig.~\ref{fig4} is
shown. With initial pulse separation $T=6$ (fig.\ref{fig4}ab) 
the initial soliton eigenvalues are 
$\lambda_1=i0.4975$ and $\lambda_2=i0.5025$. 
At initial stage of propagation in DOF, solitons are attracted (Fig.\ref{fig4}a). 
And after collision they are repel.
Due to the variation of the fiber dispersion, the collision of solitons occurs 
much earlier than predicted by the theory (\ref{z_p}). The collision point is 
located at $z=18.83\,\mbox{km}$ (Fig.~\ref{fig4}a). while analytical solution 
of unperturbed NLS equation (\ref{z_p}) predicts the collision 
at $z_c=z_p/2=31.97\,\mbox{km}$.
After collision, the solitons become propagating with the different 
group velocities (Fig.\ref{fig4}a).
Group velocity of soliton is associated with
the shift of its carrier frequency $\Delta \Omega$ (\ref{frequency}),
which is defined by real part of soliton eigenvalue $Re(\lambda_j)$.

With initial pulse separation $T=4$ (\ref{fig4}cd) the oscillation period 
is $z_p=8.66\,\mbox{km}$, initial soliton eigenvalues are 
$\lambda_1=i0.4814$ and $\lambda_2=i0.5180$. 
The soliton collision leads to the formation of central
high-intensity pulse Fig.~\ref{fig4}c. A part of radiation emerges into
dispersive wave and goes out of the central pulse. The generation of the
dispersive wave is forced by periodical perturbation of the dispersion
\cite{Hasegava}. Inverse scattering analysis show that inelastic collision
leads to fusion of solitons and formation of two-soliton breather 
(Fig.~\ref{fig4}d). The central frequencies of the solitons 
remains unchanged $(Re(\lambda_j)=0)$.
Figure \ref{fig4} show that the DOF can be used to rearrange both the solitons carrier frequencies
and the solitons energies. That reflects in the rearrangement of
eigenvalues. The shift of carrier frequency of soliton is connected
with the real part of its eigenvalue (\ref{frequency}) and the soliton amplitude
is given by imaginary part of the soliton eigenvalue  (\ref{R}).

\begin{figure} [ht]
   \begin{center}
   \begin{tabular}{cc}
   (a) & (b) \\
   \includegraphics[height=3.5cm]{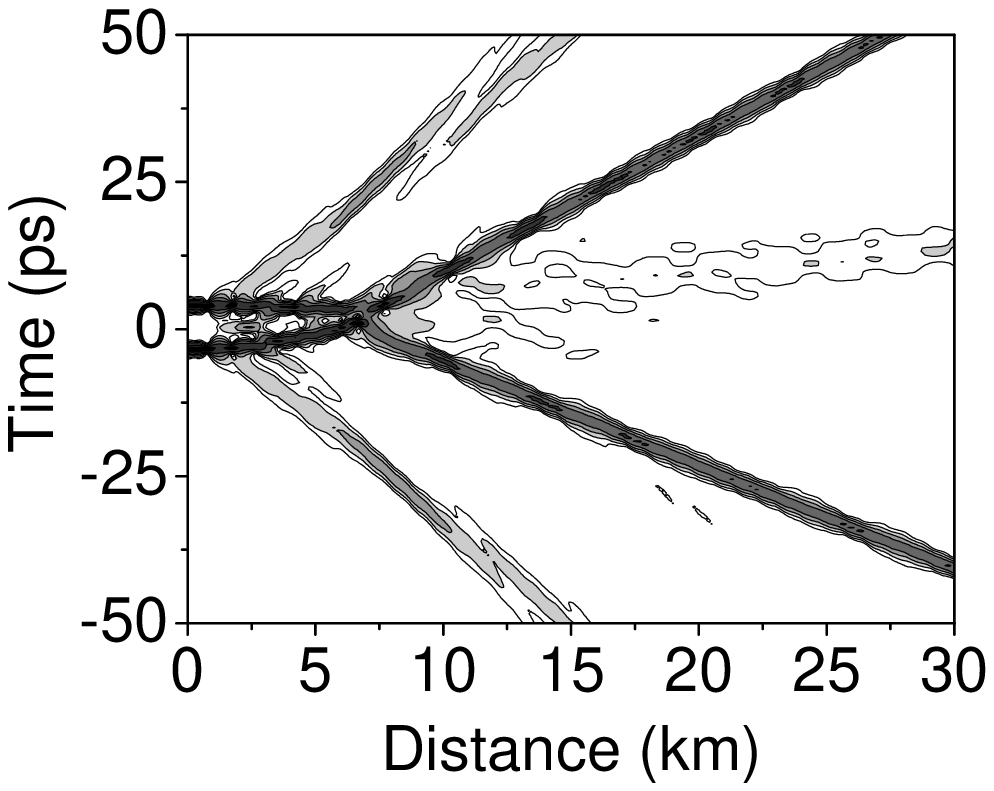} &
   \includegraphics[height=3.5cm]{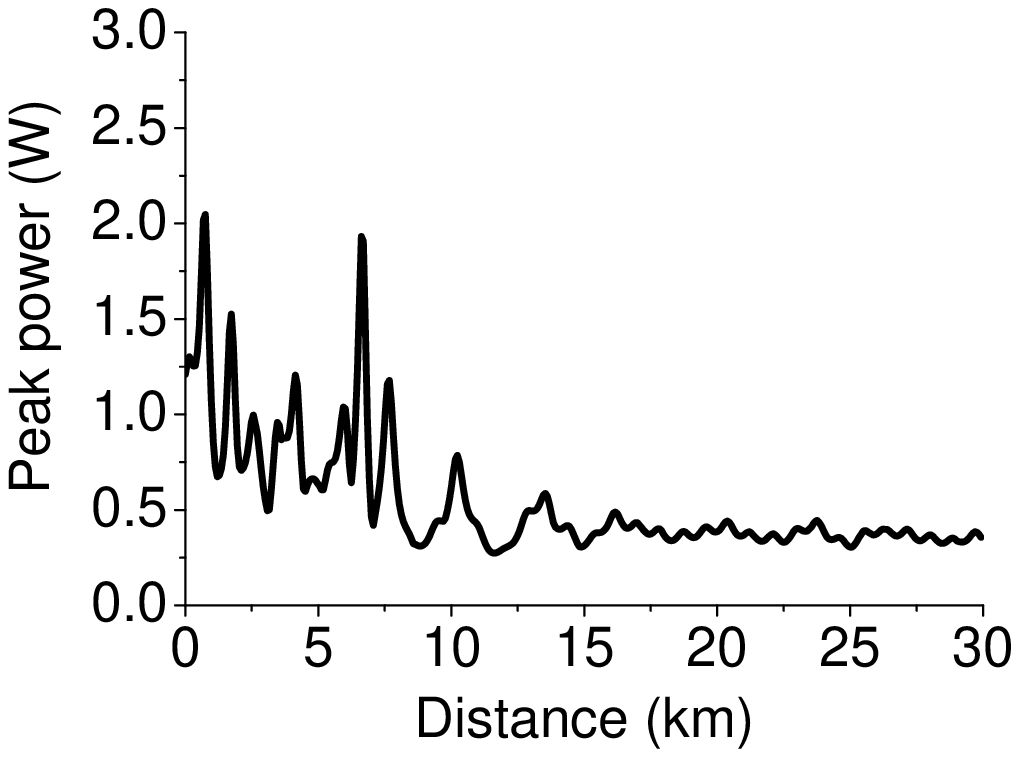} \\
   (c) & (d) \\
   \includegraphics[height=3.5cm]{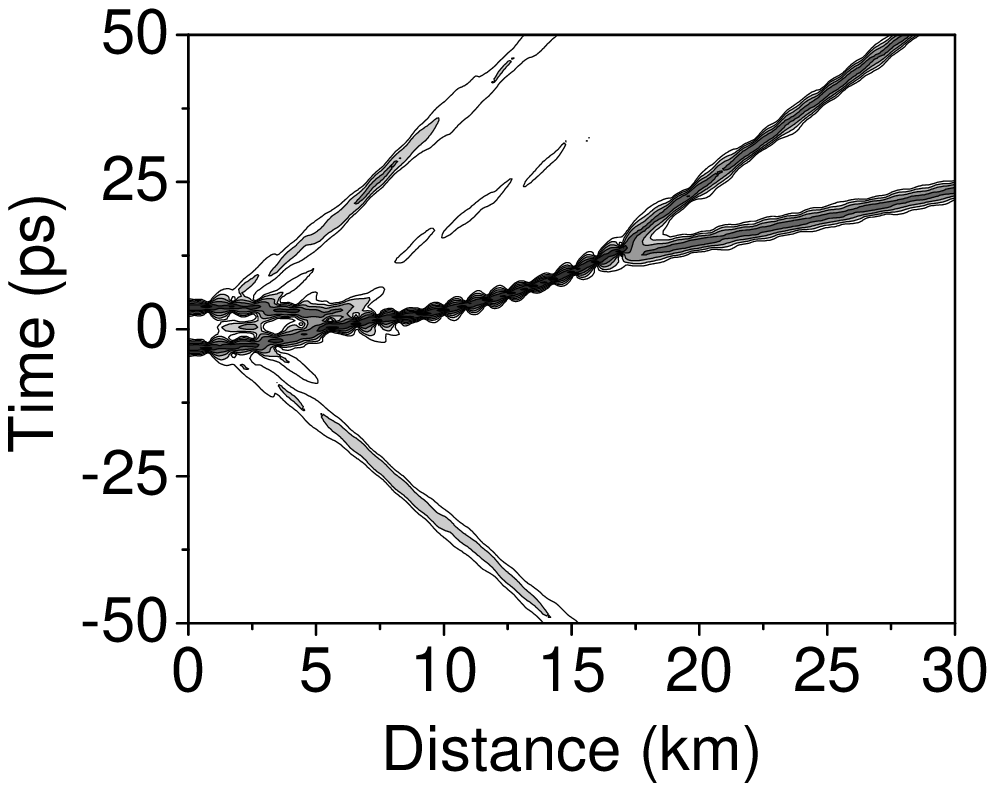} &
   \includegraphics[height=3.5cm]{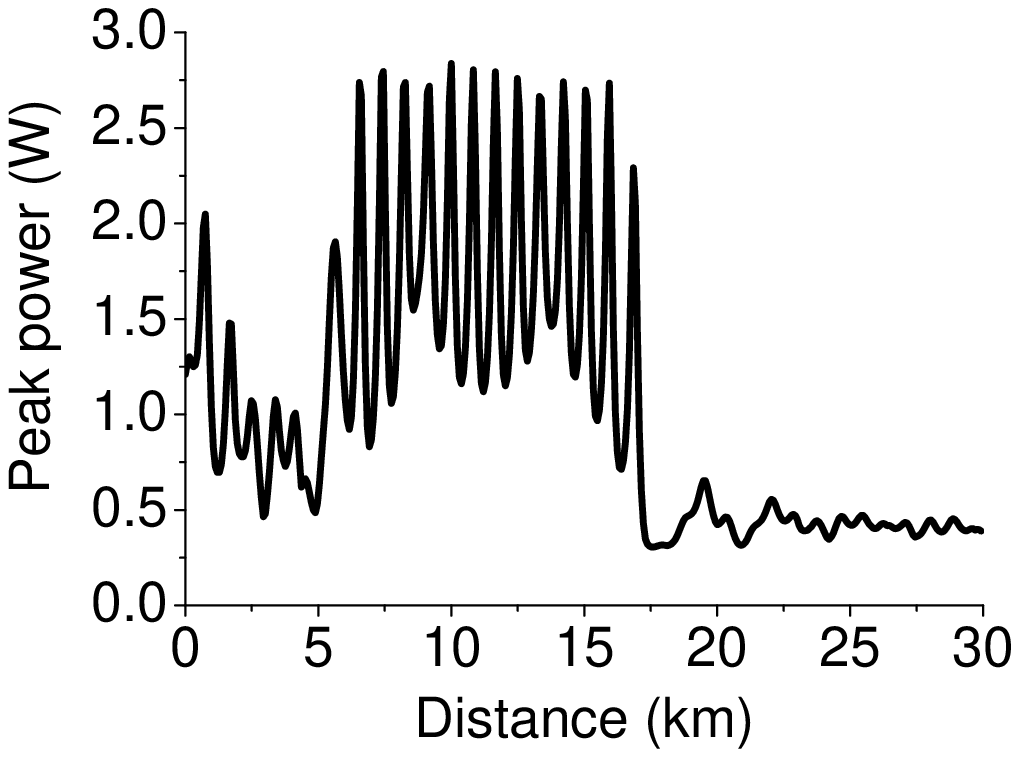}
   \end{tabular}
   \end{center}
   \caption[figure 5] 
   { \label{fig5} 
   Inelastic collision of two in-phase solitons in presence of the 
   stimulated Raman scattering and third-oder dispersion.
   {(a), (b),} $T=6.5$;  {(c), (d)} $T=6$;
   {(a), (c),} Contour plots of the field intensity $|A(z,t)|^2$; 
   {(b), (d),} Peak power.
   Modulation period is $z_m = 847.5\,\mbox{m}$, modulation phase is $\varphi_m =0$.}
   \end{figure}

At the considered parameters stimulated Raman scattering plays 
an important role in the distortion of the process of soliton collision.
Figure \ref{fig5} shows results of numerical solution
of modified nonlinear Schr\"odinger equation (\ref{schroedinger}) with
$f_R = 0.18$, $\alpha =0$, $\beta_2(z)$ and $\beta_3(z)$ are defined
by (\ref{beta2}).

The change of group velocities of two solitons 
(\ref{Aini}) with initial pulse peak separation $T=6.5$
in the Fig.~\ref{fig5} is shown. 
The initial stage of pulses propagation
is accompanied by the generation of the two dispersive
waves which quickly goes out of the computation window. At the propagation distance
$z=6.65\,\mbox{km}$ inelastic collision of initial pulses 
leads to formation of two solitons with different group velocities (Fig.~\ref{fig5}a).
The peak power of the resulting pulses 
decreases due to emergence of dispersive waves (Fig.~\ref{fig5}b).

Variation of the time separation between initial solitons allows to control
soliton dynamics. For $T=6$ (Fig.~\ref{fig5}c) initial pulse merges
into high-intensity breather at $z=7$~km. This breather
propagates up to the distance $z=17$~km. 
The peak power of breather $(7~\mbox{km}<z<17~\mbox{km})$ is two times higher
than the peak power of input pulses (Fig.~\ref{fig5}d). 
After $z=17$~km the pulse
decays into two low-intensity first-order solitons. 

During propagation of first-order solitons, only a reduction
in pulse energy due to the emission of a dispersive wave is
possible. The dispersive
wave has the highest intensity when the dispersion variation
period coincides with the soliton period \cite{Hasegava}. The period of
dispersion variation $z_m$ does not coincide with
the period of output first-order solitons (Fig.~\ref{fig5}), so the energy loss due to the emission
of a dispersive wave is a rather slow process.
The high-intensity pulses (Fig.~\ref{fig5}c) formed due to inelastic collisions 
of the solitons can propagate at the relative large propagation distances. 
However perturbation effects can lead to disappearance of high-intensity pulse.
The solitons fusion and dynamics of high-intensity pulse can be considered
as a birth and disappearance of rogue wave.

\section{Discussion and outlook}

Periodic modulation of the fiber diameter as a means of controlling soliton
interaction and breather fission was proposed. We demonstrate all
all-optical fiber technique for wavelength conversion  between picosecond
pulses based on splitting of optical breather into two pulses  with
different carrier frequencies and propagating with different group
velocities. Our experiments showed clear dependence of the time distance between output pulses
on the fiber modulation parameters, as predicted in theory.
Splitting of the breather in dispersion oscillating fiber allows to build up
pulse frequency division multiplexing from single picosecond laser source.
Frequency division multiplexing of picosecond pulses can be used
for high-speed multiwavelength optical clock for terabit hybrid  
WDM/TDM passive optical networks \cite{d1,d2}. The dispersion oscillating 
fiber can change eigenvalues that can be used for encoding channels
in nonlinear frequency division multiplexing optical fiber networks \cite{Le}.


\section*{ACKNOWLEDGMENTS}
The reported study was funded by RFBR and DST according to the research project No. 19-52-45012

\end{document}